# White Paper - Bio-Signals-based Situation Comparison Approach to Predict Pain

Uri Kartoun, Ph.D.

*Abstract*—This paper describes a time-series-based classification approach to identify similarities between bio-medical-based situations. The proposed approach allows classifying collections of time-series representing bio-medical measurements, i.e., situations, regardless of the type, the length and the quantity of the time-series a situation comprised of.

## I. INTRODUCTION

IN our era of cloud computing, big data, and wearable sensors, large number of sensors will soon continuously monitor our biological processes. Real-time sensing using wearable technologies includes variety of measurements such as heart rate, brain activity, and hydration levels to name a few. [1] provide a comprehensive review of variety of systems and sensing devices, including bio-chemical sensors for collecting fluid [2][3], accelerometers to identify epilepsy seizures [4], accelerometers to monitor *COPD* [5], photoplethysmographic bio-sensors to monitor cardiovascular activity [6]-[8], and pulse oximeters to monitor oxygen saturation [9]. Additional examples include using *Complex Event Processing* with bio-sensors, *RFID*, and accelerometers to evaluate health-related events [10], *Zigbee* modules to monitor temperature and heart rate [11], *EMG*, *EDA*, and pressure respiration sensors to monitor mental stress [12], and *ECG* recording to measure respiration rate [13].

The structure of the paper is as follows: Section II details the definitions of the time-series classification approach. Section III describes a medical use case to predict an upcoming wave of physical pain. Conclusions are provided in Section IV.

## II. TIME-SERIES-BASED SITUATION COMPARISON

A situation is defined as a collection of one or more events that occur over a range of time. Examples include high-impact occurrences such as an earthquake or a flood, or a situation could be a planned occurrence such as an airplane taking off, or a vehicle crossing an intersection. An example for a situation could also be the occurrence of heart attack or epilepsy seizure, cyber-attack, or it could be a stock market crisis such as the *2010 Flash Crash* [14].

A *bio-medical baseline situation (BMBS)* is defined as a collection of bio-medical time-series that were measured prior to a significant event (e.g., a heart attack or an epilepsy seizure). A sampled situation denoted as *bio-medical situation of interest (BMSI)*, if determined as significantly similar to the *BMBS* could imply upon the upcoming similar significant event. The ability to evaluate the level of similarity between two situations by identifying similar inter-patterns between the situations has the potential to alert upon undesired situations, and allows taking preventive actions in advance (Fig. 1).

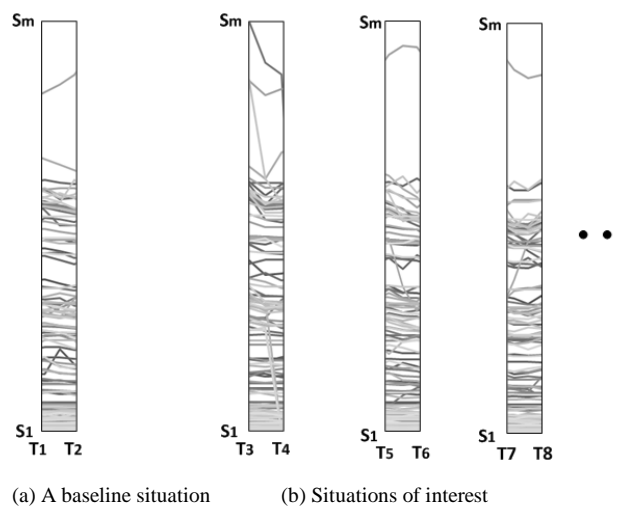

(a) A baseline situation      (b) Situations of interest

Fig. 1. An example for time-series-based bio-medical situations.

As shown in Fig. 1(a), a baseline situation comprised of $m$ time-series $S_1, S_2, S_i, \ldots S_m$ sampled from time $T_1$ to time $T_2$. A situation of interest comprised of the same type of $m$ time-series sampled during $T_3$ to $T_4$, etc. (Fig. 1(b)). For each situation several samples are taken. For example, for the baseline situation, sampling of time-series includes a first sampling of the values of $m$ time-series measured at time $T_1$, several more samples of the values of the $m$ time-series (e.g., every ten seconds), and a last sampling of the values of the $m$ time-series measured at time $T_2$. A similar sampling procedure is applied on each situation of interest from $T_3$ to $T_4$, from $T_5$ to $T_6$, etc. The number of samples, $n$, for a situation is determined in advance as well as the interval between samples. A componential method, such as artificial neural network, support vector machines, or decision tree learning could be used to classify the time-series in each situation and to determine the level of similarity between the two situations.

## III. Pain-Prediction Use Case

One potential medical implementation based on the proposed approach is a bio-signals-based situation comparison system to predict pain; after a surgery, a patient staying at the hospital often suffers from physical pain. The implementation could be a pain predicting system which is based on collecting and analyzing a large number of bio-signal readings the patient's body produces measured in real-time. The implementation will be based on dynamically comparing human situations. A situation is a collection of time-series representing bio-signals measured during a pre-defined time-range (for example, 15 minutes). A *BMBS* is determined as a known situation before which the patient experiments a subjectively measured significant level of pain. The implementation may include a method that dynamically measures current bio-signals-based situations and compares them to the BMBS. A similarity rank given in percent between the two situations determines how close a current situation, i.e., a *BMSI*, is to the *BMBS* that was associated with an upcoming wave of pain. Such an implementation has the potential to notify in advance that the patient is expected to experiment a significant wave of pain. This way, the medical staff would provide the patient with the appropriate pain killers long before the pain starts to affect - the patient will suffer less, and more likely will stay less at the hospital.

## IV. Conclusions

The approach described in the paper could be implemented on a large collection of bio-medical sensors to allow sampling and classifying unspecified number of time-series measurements. Additional sensor readings, could be acquired without modifying the design and without having to develop additional software components. Another advantage of the approach is that it has the potential to reduce the complexity that involves in fusioning the measurements acquired by multiple sensors. The ability to fusion readings from multiple and unspecified number of sensors, including new types of sensors, has the potential to include additional sensors as part of an existing monitoring bio-medical system (e.g., *Q-Sensor* [15]), and may include in the future sensors that not yet fully been explored. Examples for futuristic sensors include physical-measuring wearable sensors that measure the level of dehydration [16], sensors that measure the level of exposure to sunlight, and sensors that measure weight and height. Examples for futuristic sensors include also psychological-measuring wearable sensors that measure the level of patience, sensors that measure the level of boredom, and sensors that measure the level of fatigue.